\documentclass[twocolumn]{aastex631}

\begin{document}

\title{Organic chemistry in the H$_2$-bearing, CO-rich interstellar ice layer at temperatures relevant to dense cloud interiors}

\author[0000-0001-6496-9791]{Rafael Mart\'in-Dom\'enech}
\altaffiliation{The current affililation is Centro de Astrobiolog\'ia (CSIC-INTA), but the reported experiments were performed during the previous appointment at Center for Astrophysics $|$ Harvard \& Smithsonian.}
\affil{Centro de Astrobiolog\'ia (CSIC-INTA)\\
Carretera de Ajalvir, km. 4, Torrej\'on de Ardoz, E-28850, Madrid, Spain}
\correspondingauthor{Rafael Mart\'in-Dom\'enech}
\email{rmartin@cab.inta-csic.es}

\author[0000-0003-2322-7526]{Alexander DelFranco}
\affil{Center for Astrophysics $|$ Harvard \& Smithsonian\\
60 Garden St., Cambridge, MA 02138, USA}
\affil{Physics and Astronomy Department, Amherst College, \\
21 Merrill Science Drive, \\
Amherst, MA 01002, USA}

\author[0000-0001-8798-1347]{Karin I. \"Oberg}
\affil{Center for Astrophysics $|$ Harvard \& Smithsonian\\
60 Garden St., Cambridge, MA 02138, USA}

\author[0000-0003-2761-4312]{Mahesh Rajappan}
\affil{Center for Astrophysics $|$ Harvard \& Smithsonian\\
60 Garden St., Cambridge, MA 02138, USA}

\begin{abstract}

Ice chemistry in the dense, cold interstellar medium (ISM) is probably responsible for the formation of interstellar complex organic molecules (COMs). 
Recent laboratory experiments performed at T $\sim$ 4 K have shown that irradiation of CO:N$_2$ ice samples analog to the CO-rich interstellar ice layer can contribute to the formation of COMs when H$_2$ molecules are present. 
We have tested this organic chemistry under a broader range of conditions relevant to the interior of dense clouds by irradiating CO:$^{15}$N$_2$:H$_2$ ice samples with 2 keV electrons in the 4$-$15 K temperature range. 
The H$_2$ ice abundance depended on both, the ice formation temperature and the thermal evolution of the samples.    
Formation of H-bearing organics such as formaldehyde (H$_2$CO), ketene (C$_2$H$_2$O), and isocyanic acid (H$^{15}$NCO) was observed upon irradiation of ice samples formed at temperatures up to 10 K,  
and also in ices formed at 6 K and subsequently warmed up and irradiated at temperatures up to 15 K. 
These results suggest that a fraction of the H$_2$ molecules in dense cloud interiors might be entrapped in the CO-rich layer of interstellar ice mantles, and  
that energetic processing of this layer could entail an additional contribution to the formation of COMs in the coldest regions of the ISM. 

\end{abstract}


\section{Introduction} \label{sec:intro}

Interstellar complex organic molecules (COMs or iCOMs in the literature) are potential precursors of yet more complex prebiotic species, and could have played a role in the origin of life on Earth. 
COMs are thought to arise from chemical processes taking place in ice mantles present in the coldest regions of the interstellar medium \citep[ISM,][]{garrod13}. 
These ice mantles are formed on the surface of dust grains in the interior of dense molecular clouds, with temperatures down to T $\sim$ 7 K and densities above n $\geq$ 10$^4$ cm$^{-3}$ \citep{boogert15}.  
COMs are eventually incorporated into protostellar and protoplanetary disks \citep[see, e.g.][]{alice21} during the star formation process, and have also been detected in Solar System comets \citep[see, e.g.,][]{goesmann15,altwegg15}. 

Interstellar ice chemistry usually involves radical-radical or radical-neutral reactions. 
Ice radicals can be formed through multiple pathways, including atom-addition reactions,  H abstraction, or energetic processing of the ice molecules with dissociative radiation. 
Even though the interior of dense clouds are shielded from the interstellar UV field, a secondary UV field along with secondary keV electrons and suprathermal particles generated via Coulomb explosion are available to interact with ice molecules thanks to the interaction of cosmic rays with the gas-phase H$_2$ molecules \citep{cecchi92,shen04} or the ice molecules themselves \citep{bennett05}, thus enabling the energetic processing of the ice mantles in the interior of dense clouds.  
%
Ice chemistry induced by energetic processing can proceed not only in H$_2$O-dominated ices, 
but also in the CO-rich ice layer covering the H$_2$O-rich mantles in regions where the temperature drops below $\sim$20 K and episodes of so-called "catastrophic" CO freeze-out take place \citep{boogert15}. 

The CO-rich ice layer probably contains other volatile species accreted from the gas phase such as N$_2$ and O$_2$ \citep{boogert15}. 
Some authors have also speculated about the presence of H$_2$ molecules in the CO-rich layer \citep[see, e.g.,][]{chuang18}. 
\citet{martin20} recently showed that irradiation of CO:$^{15}$N$_2$:H$_2$ ice analogs at $\sim$4 K with 2 keV electrons and Ly$\alpha$ photons leads to the formation of ice chemistry products spanning a range of complexities. 
The identified products included simple species such as CO$_2$, C$_2$O (and other carbon chain oxides) and CH$_4$, as well as organics such as H$_2$CO, C$_2$H$_2$O, and H$^{15}$NCO. We note that the latter is particularly interesting from a prebiotic chemistry perspective since it contains the amide bond ($-$(H$-$)N$-$C($=$O)$-$) that links amino acids into proteins. 
These results suggested that the energetic processing of the CO-rich ice layer could also contribute to the formation of COMs in the dense ISM, in addition to other more often studied formation pathways such as the the chemistry induced by energetic processing of the H$_2$O-rich layer
\citep[see, e.g., the review in][]{obergrev}, or the H-atom addition reactions in the CO-rich layer \citep{watanabe02,watanabe08,fuchs09,hama13,linnartz15,fedoseev15a,fedoseev17,chuang16,chuang17,qasim19a,qasim19b}.
In the case of some species such as HNCO, whose observed abundances might not be fully explained by previously proposed formation pathways \citep[see][and references therein]{martin20}, energetic processing of the CO-rich ice layer could represent a significant contribution to their interstellar budgets.  

Prior to the experiments reported in \citet{martin20}, \citet{chuang18} had reported the formation of HCO$^.$ radicals and H$_2$CO molecules upon UV irradiation of CO:H$_2$ ice samples at temperatures up to 20 K, with a strong temperature dependence for the HCO$^.$ formation yield (higher temperatures leading to lower yields). 
In this work, we have explored in the laboratory the chemistry induced by 2 keV electron irradiation of CO:$^{15}$N$_2$:H$_2$ ice analogs at a range of temperatures relevant to the interior of dense molecular clouds (4$-$15 K), in order to 
i) test whether the reported organic chemistry at $\sim$4 K takes place at temperatures relevant to the dense cloud interiors; and ii) constrain how this chemistry is affected by the temperature and abundance of H$_2$ molecules in the ice. 
%
The experimental setup used in these laboratory experiments is described in Sect. \ref{sec:exp}. The results are presented in Sect. \ref{sec:results} and discussed in Sect. \ref{sec:disc}. Finally, the main conclusions are listed in Sect. \ref{sec:conc}.

\section{Methods} \label{sec:exp}

\begin{deluxetable*}{cccccccccc}
\tablecaption{Summary of the experiments simulating the energetic processing of the CO-rich ice layer.\label{table_exp}}
\tablehead{
\colhead{Exp.} & \colhead{Ice comp.} & \multicolumn{2}{c}{Composition ratio$^a$} & \multicolumn{2}{c}{N(CO) (ML)$^{b}$}  &  \colhead{H$_2$ IR int. abs.$^c$} & \multicolumn{2}{c}{Temp. (K)} & \colhead{Irrad. energy }\\
&& \colhead{gas mixture} & \colhead{ice sample} & \colhead{Initial} & \colhead{Active$^d$} & & \colhead{Dosing} & \colhead{Irrad.} & \colhead{($\times$ 10$^{18}$ eV)}}
\startdata
1 & CO:$^{15}$N$_2$:H$_2$ & 1.0:1.0:3.1 & 1.5:1.5:1 & 460 & 140 & 0.0128 $\pm$ 0.0004 & 4.7 & 4.7 & 1.8 \\ 
2 & CO:$^{15}$N$_2$:H$_2$ & 1.0:1.0:3.1 & 10:10:1 & 390 & 135 & 0.0014 $\pm$ 0.0003 & 8.0 & 8.0 & 1.7 \\ 
3 & CO:$^{15}$N$_2$:H$_2$ & 1.0:1.1:3.6 & 30:30:$\le$1 & 320 & 140 & $\le$0.0004$^e$ & 10.0 & 10.0 & 2.0 \\ 
\hline
4 & CO:$^{15}$N$_2$:H$_2$ & 1.0:1.1:3.6 & 4:4:1 & 360 & 125 & 0.0033 $\pm$ 0.0003 & 6.0 & 7.0 & 1.8 \\ 
5 & CO:$^{15}$N$_2$:H$_2$ & 1.0:1.1:3.8 & 4:4:1 & 430 & 125 & 0.0042 $\pm$ 0.0003 & 5.9 & 9.0 & 1.8 \\ 
6 & CO:$^{15}$N$_2$:H$_2$ & 1.0:1.0:3.7 & 10:10:1 & 410 & 135 & 0.0015 $\pm$ 0.0003 & 6.0 & 15.0 & 1.9 \\ 
\enddata
\tablecomments{
$^a$ The composition ratios of the gas mixtures used to form the ice samples were measured with a QMS (Appendix \ref{sec:gas_composition_app}).  
The composition ratios of the ice samples are a rough estimation using the corresponding H$_2$ IR integrated absorbance, and assuming that the H$_2$ sticking coefficient in Exp. 1 was 0.22 (see Sect. \ref{sec:ice}). 
$^{b}$ CO column density in monolayers (1 ML = 10$^{15}$ molecules cm$^{-2}$). A 30\% uncertainty was assumed, due to the estimated uncertainty for the 4253 cm$^{-1}$ CO IR band strength in reflection-absorption mode (Appendix \ref{sec:ir_app}).
$^c$The integrated IR absorbance and corresponding uncertainty are obtained from a Gaussian fit to the feature shown in Fig. \ref{fig:ir_series1_initial}.
$^d$ Chemically active CO column density corresponding to a total electron penetration depth of 370 ML in Exp. 1, and 285 ML in Experiments 2$-$6 (Sect. \ref{sec:e-}). 
$^e$ 3$\sigma$ upper limit.} 
\end{deluxetable*}{}

Table \ref{table_exp} summarizes the laboratory experiments carried out for this paper. 
Two series of experiments were performed. 
In Experiments 1$-$3, CO:$^{15}$N$_2$:H$_2$ ice samples were deposited and irradiated with 2 keV electrons at a certain temperature (4.7, 8.0, and 10.0 K, respectively). 
In Experiments 4$-$6, the samples were deposited at $\sim$6 K and irradiated at a higher temperature (7.0, 9.0, and 15.0 K, respectively). 

The laboratory experiments were carried out in the SPACE TIGER\footnote{Surface Photoprocessing Apparatus Creating Experiments To Investigate Grain Energetic Reactions} experimental setup. 
SPACE TIGER consists of a 500 mm diameter stainless steel, ultra-high-vacuum (UHV) chamber (custom made by  Pfeiffer Vacuum) where a base pressure of $\sim$2 $\times$ 10$^{-10}$ Torr at room temperature is reached thanks to the combination of magnetically levitated turbomolecular and scroll pumps. 
This setup is described in detail in \citet{pavlo22}.  
The relevant features used for this work are presented below. 

\subsection{Ice sample preparation}\label{sec:ice}

The CO:$^{15}$N$_2$:H$_2$ ice samples used in Experiments 1$-$6 (Table \ref{table_exp}) were grown on a 12 mm diameter copper substrate, mounted on an oxygen-free high thermal conductivity (OFHC) copper sample holder located at the center of the UHV chamber.  
The sample holder is connected to the second stage of a closed-cycle He cryostat (Model DE210B-g, Advance Research Systems, Inc.) powered by an ARS-10HW compressor, and can be cooled down to $\sim$4 K. 
The substrate temperature is monitored with a calibrated Si diode sensor located close to the copper substrate, with a 
0.1 K relative uncertainty. 
The combination of the cryostat with a 50 W  silicon nitride cartridge heater rod (Bach Resistor Ceramics) allows to set the temperature of the substrate in the 4$-$300 K range using a Lakeshore Model 336 temperature controller. 
The dosing temperature in Experiments 1$-$6 is indicated in the seventh column of Table \ref{table_exp}. 
The temperature of the substrate remained stable within the 0.1 K accuracy during dosing and irradiation of the ice samples.
Ice samples deposited in Experiments 1$-$6 were expected to be amorphous \citep{kouchi21a,kouchi21b}.

The ice samples were formed by exposing the substrate to a gas mixture with the desired composition prepared in a independently pumped gas line assembly. 
The gas mixtures were composed by CO (gas, 99.95\%, Aldrich), $^{15}$N$_2$ (gas, 98\%, Aldrich), and H$_2$ (gas, $>$99.99\% purity, Aldrich). 
We used a similar dosing pressure and time in all experiments, that led to an ice deposition rate of, approximately, 5$-$8 ML/s in Experiments 1$-$6 (1 ML = 10$^{15}$ molecules cm$^{-2}$). 
The gas mixtures were introduced in the UHV chamber through a 10 mm diameter dosing pipe in close proximity ($<$2 mm) to the cold substrate.  We thus assumed a 10 mm diameter size for the deposited ice samples. 
The gas mixture composition was measured in the UHV chamber before ice deposition using a quadrupole mass spectrometer (QMS, Sect. \ref{sec:tpd}). 
The measured QMS signals corresponding to CO, $^{15}$N$_2$ and H$_2$ were transformed into partial pressures with a conversion factor previously calculated for the pure gases (see Appendix \ref{sec:gas_composition_app}). 
The measured composition ratios are indicated in the third column of Table \ref{table_exp}.

We note that the H$_2$ abundance in the gas mixture should be considered an upper limit for the H$_2$ abundance in the ice sample, since the sticking coefficient of H$_2$ molecules onto a substrate is expected to be lower than unity,  
whereas a $\sim$1 sticking coefficient is generally assumed for most of the species, including CO and $^{15}$N$_2$. 
%
In this work we assumed, as a first approximation, a H$_2$ sticking coefficient of 0.22 for Exp. 1 (dosing temperature of $\sim$ 4.7 K), as reported in \citet{german21} for room-temperature H$_2$ molecules sticking on a CO-ice surface at 5 K.   
We then used 
this experiment as a reference 
to estimate the H$_2$ ice abundance in the experiments with higher dosing temperatures (Sect. \ref{sec:h2}). 
The resulting estimated ice composition is indicated in the fourth column of Table \ref{table_exp}.

\subsection{IR ice spectroscopy}\label{sec:ir}
The ice samples in Experiments 1$-$6 were monitored through reflection-absorption infrared spectroscopy (RAIRS) using a Bruker 70v Fourier transform infrared (FTIR) spectrometer and a liquid-nitrogen-cooled MCT detector. 
The IR beam had an incidence angle of 70$^\circ$ with respect to the substrate surface normal, resulting in an elliptical beam at the sample position, with a 14.6 mm major axis and a 5 mm minor axis.  
The spectra were averaged over 512 interferograms, and collected with a resolution of 1 cm$^{-1}$ in the 5000$-$800 cm$^{-1}$ range. 

\begin{table*}[ht!]
\centering
\caption{Band strengths of selected features in pure ice IR spectra collected in transmission mode (unless otherwise indicated). A  20\% uncertainty is generally assumed for the band strength values reported in the literature \citep[see, e.g.,][]{dhendecourt86}. 
\label{table_band}}
\begin{tabular}{cccc}
\hline
\hline
Molecule&Wavenumber&Band strength&Reference\\
&(cm$^{-1}$)&(cm molec$^{-1}$)&\\
\hline
CO & 2139 & 1.1 $\times 10^{-17}$ & \citet{gerakines95}\\
CO & 4253 & 7.9 $\times 10^{-20}$ & This work\\
CO & 4253 & 1.8 $\times 10^{-19}$$^a$ & This work\\
CO$_2$ & 2343 & 7.6 $\times 10^{-17}$& \citet{gerakines95}\\
C$_2$O & 1989 & 1 $\times 10^{-17}$& \citet{palumbo08}\\
CH$_{4}$ &1304 & 6.4 $\times 10^{-18}$& \citet{mulas98}\\
H$_2$CO & 1494 & 3.9 $\times 10^{-18}$& \citet{schutte93}\\
\hline
\end{tabular}
\tablecomments{$^{a}$Measured in reflection-absorption mode. A 30\% uncertainty was assumed for this value (Appendix \ref{sec:ir_app}).}
\end{table*}

For those species with detected IR features, the corresponding ice column densities can be calculated from the integrated IR absorbances, following the equation:

\begin{equation}
N(X)=\frac{1}{A_X}\int_{band}{\tau_{\nu} \ d\nu},
\label{eqn}
\end{equation}

\noindent where $N(X)$ is the column density of species $X$ in molecules cm$^{-2}$, $\tau_{\nu}$ is the optical depth of the absorption band (2.3 times the absorbance), and $A_X$ is the band strength of the IR feature in cm molecule$^{-1}$. 

The band strengths $A_X$ of the IR features observed in reflection-absorption IR spectroscopy differ from those reported for the IR spectra collected in transmission mode, and are setup-specific \citep[see, e.g.,][]{chuang18}.   
As part of this work, we calculated the band strength of the CO IR feature observed at 4253 cm$^{-1}$ in reflection-absorption mode for the SPACE TIGER setup (Appendix \ref{sec:ir_app}).  
We used the weaker 4253 cm$^{-1}$ IR feature instead of the main 2139 cm$^{-1}$ band because the IR absorbance in reflection-absorption mode shows a non-linear behavior with the species column density (making Eq. \ref{eqn} no longer valid) above a certain threshold ice thickness \citep{oberg09}, with lower threshold ice thicknesses for stronger IR bands. 
The estimated value for the CO 4253 cm$^{-1}$ IR feature in reflection-absorption mode is listed in Table \ref{table_band}, and the resulting initial CO ice column density in Experiments 1$-$6 is indicated in the fifth column of Table \ref{table_exp}.

\subsection{2 keV electron irradiation of the ice samples}\label{sec:e-}

The deposited ice samples were electron irradiated using a ELG-2/EGPS-1022 low energy electron source system (Kimball Physics) with an electron energy of 2 keV.  
The keV electron bombardment of the ice samples mimicked the effect of the so-called $\delta$-electrons (with energies of a few keV) generated by the cosmic rays (mainly protons with a distribution maximum of $\sim$10 MeV) through electronic energy transfer to the ice molecules in the interior of dense clouds \citep{bennett05}.
The electron beam had an incidence angle of 34$^\circ$, resulting in an elliptical spot at the sample position, with a 10 mm minor axis (i.e., the ice sample diameter) and a 12 mm major axis \citep[measured with a phosphor screen,][]{pavlo22}.
The average electron beam current was $\sim$65 nA, and the sample irradiation time was 45$-$50 min, leading to a total incident energy of $\sim$1.8 $\times$ 10$^{18}$ eV (last column of Table \ref{table_exp}). 
The total incident energy fluence of $\sim$2.3 $\times$ 10$^{18}$ eV cm$^{-2}$ in Experiments 1$-$6 is similar to the estimated energy fluence deposited by the cosmic rays into the interstellar ice mantles present in the interior of dense clouds during $\sim$2 $\times$ 10$^6$ yr  \citep[see][and ref. therein]{jones11}, on the order of the typical lifetime of a dense cloud. 

The 2 keV electron penetration depth depends on the ice composition and density. 
%
%
%
%
In this work, we assumed a penetration depth of 370 ML for Exp. 1 (i.e., 95\% of the electron energy was lost in the top 370 ML), and 285 ML for the H$_2$-poor ice samples in Experiments 2$-$6. 
The penetration depths were calculated for a 1.5:1.5:1 and a 1:1:0
CO:$^{15}$N$_2$:H$_2$ ice sample, respectively, with the CASINO v2.42 code \citep{drouin07}. 
We defined the chemically active fraction of the ice samples as the ratio between the estimated electron penetration depth and the total ice thickness \citep[as in][]{martin20}.
Note that only a fraction of, approximately, 30-40\% of the ice sample was thus chemically active in Experiments 1$-$6 .  
The chemically active CO ice column density (calculated as the product of the chemically active fraction and the initial CO ice column density)is indicated in the sixth column of Table \ref{table_exp}. 

\subsection{Temperature Programmed Desorption of the irradiated ice samples}\label{sec:tpd}

\begin{figure*}
    \centering
    \gridline{
    \fig{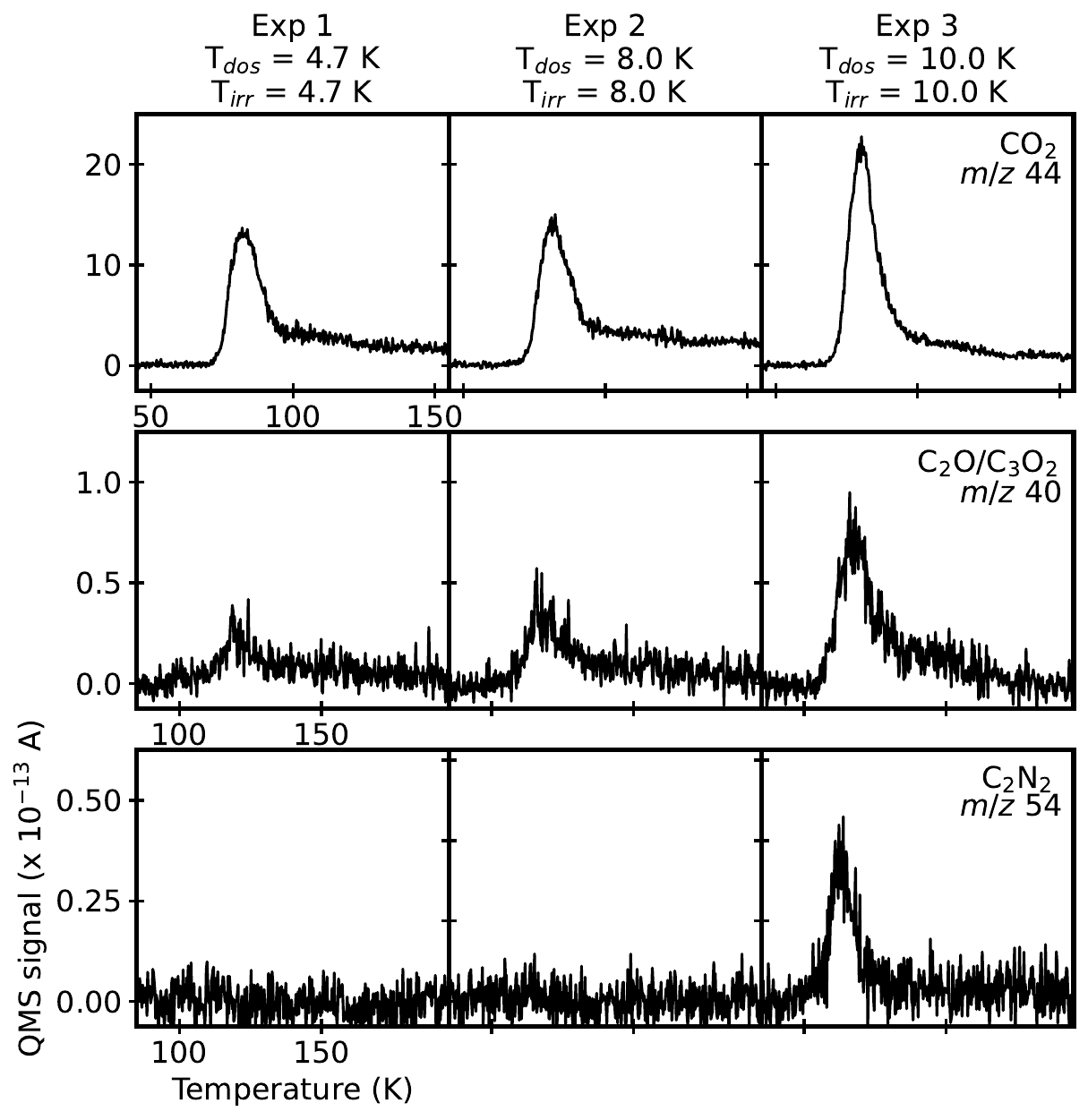}{0.49\textwidth}{}
    \fig{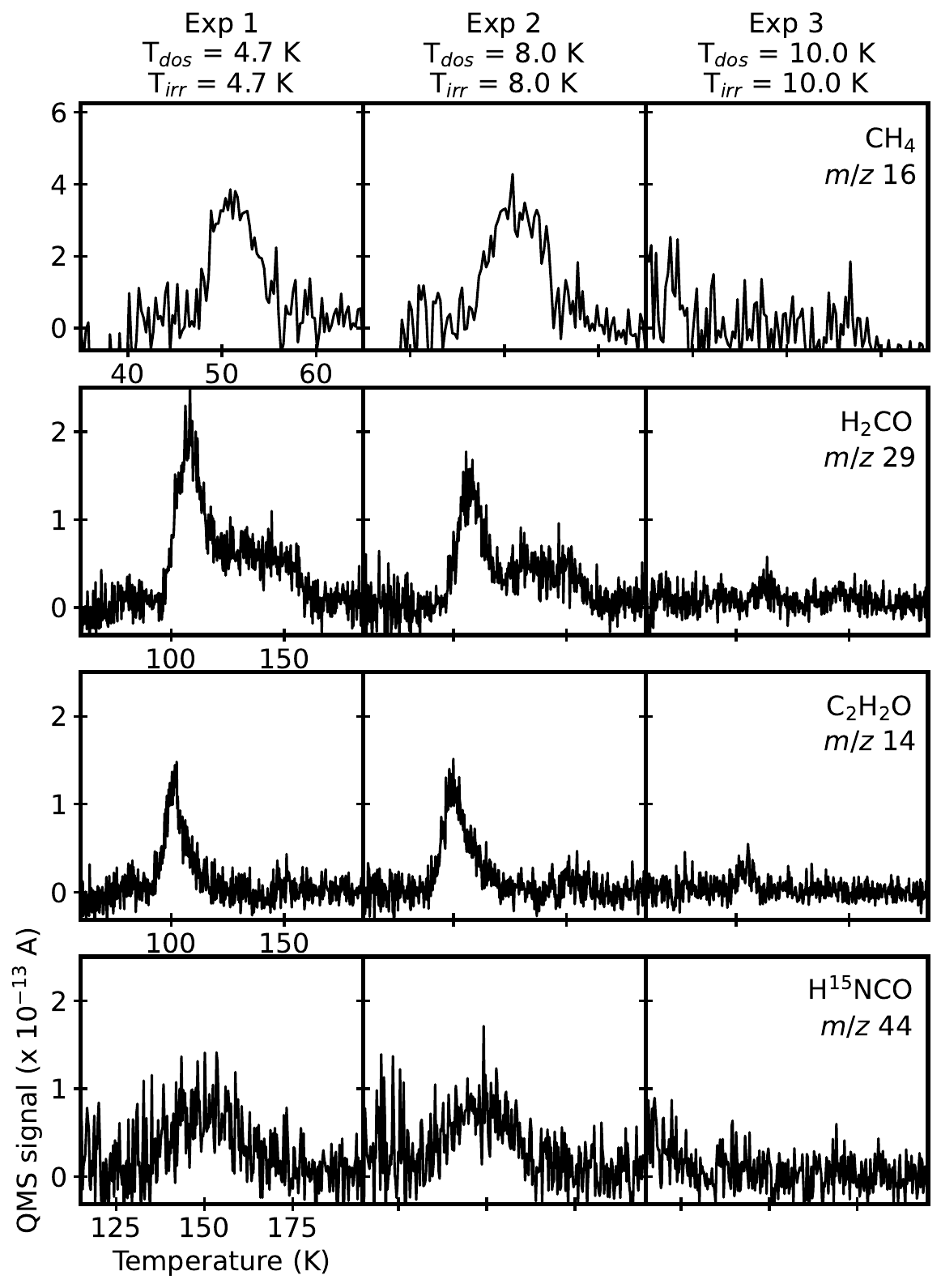}{0.475\textwidth}{}}
    \caption{TPD curves of the CO$_2$, C$_2$O/C$_3$O$_2$, and C$_2^{15}$N$_2$ main mass fragments (\textit{Left}, from top to bottom: $m/z$ 44, $m/z$ 40, and $m/z$ 54, respectively) and the CH$_4$, H$_2$CO, C$_2$H$_2$O, and H$^{15}$NCO  main mass fragments (\textit{Right}, from top to bottom: $m/z$ 16, $m/z$ 29, $m/z$ 14, and $m/z$ 44, respectively) during warm-up of CO:$^{15}$N$_2$:H$_2$ ice samples after irradiation of a similar incident energy with 2 keV electrons in Experiments 1$-$3 (from left to right).
    The TPD curves were baseline-corrected using a spline function with the \texttt{IDL} software. 
    }    \label{fig:tpd_series1}
\end{figure*}

Following the electron irradiation of the ice samples, these were warmed from the irradiation temperature (ninth column of Table \ref{table_exp}) up to 200 K with a controlled heating rate of 2 K min$^{-1}$ in a so-called temperature programmed desorption (TPD). 
The desorbing molecules were detected with a QMG 220M1 QMS (Pfeiffer, mass range 1$-$100 amu, resolution of 0.5 amu) located at $\sim$13 cm from the substrate. 
The different ice species were monitored through the main mass fragments reported in the online database of the National Institute of Standards and Technologies (NIST), incicated in Figures \ref{fig:tpd_series1} and \ref{fig:tpd_series2}. 

Calibration of the QMS (Appendix \ref{sec:qms_calib_app}) allowed an alternative estimation of the final product column density from the area under the corresponding TPD curve ($A(m/z)$) using Eq. \ref{eqn_qms}. 
This was particularly useful for those species with no detected IR features. 

\begin{figure*}[ht!]
    \centering
    \gridline{
    \fig{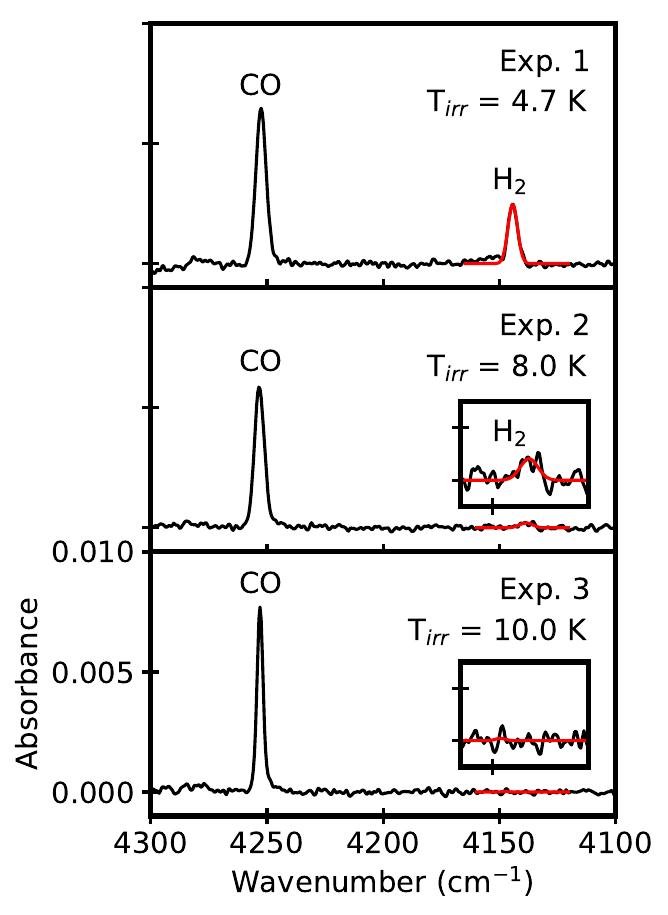}{0.3\textwidth}{}
    \fig{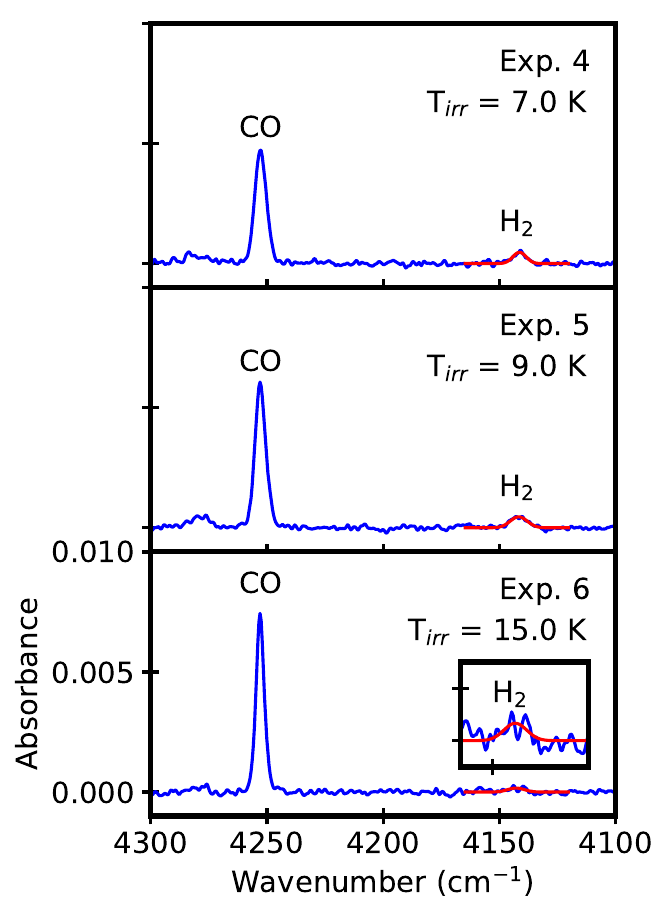}{0.3\textwidth}{}}
    \vspace{-.75cm}
    \caption{IR spectra of the initial ice samples in Experiments 1$-$3 (left panels) and Experiments 4$-$6 (right panels) collected in reflection-absorption mode at the irradiation temperature. 
    For Experiments 1$-$3 the dosing and irradiation temperatures were the same, while in Experiments 4$-$6 the dosing temperature was $\sim$6 K in all cases (see Table \ref{table_exp}). 
    Insets show a zoom-in of the region where the H$_2$ IR feature is expected.
    The Gaussian fit of the H$_2$ IR feature is shown in red in every panel.}    
    \label{fig:ir_series1_initial}
\end{figure*}{}

\subsection{Conversion yields}\label{sec:conv_methods}
In order to quantify the conversion of CO molecules into C-bearing products upon electron irradiation of the ice samples in Experiments 1$-$6, we calculated the product conversion yields as the ratio between the final product column density and the 
chemically active CO column density listed in Table \ref{table_exp} ($N_f(X)/N(CO)$). 
%
%

For those ice chemistry products with detected IR features (i.e., CO$_2$, C$_2$O, CH$_4$, and H$_2$CO), the final column density was calculated with Eq. \ref{eqn}.
A Gaussian fitting of the corresponding IR features was performed using the \texttt{curve$\_$fit} function in the SciPy library in order to calculate the IR integrated absorbance, except for CO$_2$, for which a numerical integration was performed using the \texttt{integrate.simps} function. 
The band strengths of selected features in pure ice IR spectra collected in transmission mode are listed in Table \ref{table_band}. 
In order to use Eq. \ref{eqn} with the integrated absorbances measured in reflection-absorption mode, we assumed that the relative band strengths with respect to the CO IR feature detected at 4253 cm$^{-1}$ were the same in the spectra collected in transmission and reflection-absorption mode \citep[similar to previous studies, see, e.g.,][]{oberg09, chuang18}.
In addition to the systematic 20\% uncertainty generally assumed for the reported IR band strengths (Table \ref{table_band}), and the 30\% uncertainty estimated for the 4253 cm$^{-1}$ CO IR band strength in reflection-absorption mode (Appendix \ref{sec:ir_app}), we assumed an additional 25\% experimental uncertainty that accounted for differences found in the conversion yields of experiments that should present the same results \citep{martin20}. 
This led to a total uncertainty of 45\% in the product conversion yields calculated from the integrated IR absorbances. 
We note that only the 25\% experimental uncertainty should be taken into account when evaluating differences in the yields of the same species across different experiments. 

For those products with no detected IR features (i.e., C$_2^{15}$N$_2$, C$_2$H$_2$O, and H$^{15}$NCO), the final column density was calculated from the area under the corresponding TPD curve ($A(m/z)$) using Eq. \ref{eqn_qms}. 
The TPD curves were numerically integrated with the \texttt{integrate.simps} function in SciPy. 
The proportionality constant $k_{CO}$ was estimated as part of this work (Appendix \ref{sec:qms_calib_app}), and the rest of required parameters are listed in Table \ref{table_qms}. 
In this case, we assumed a systematic 50\% uncertainty in the estimated product column densities, since some of the used parameter were rough approximations (see Table \ref{table_qms}). Along with the 30\% uncertainty in the intial CO ice column density, and the 25\% experimental uncertainty, this led to a total uncertainty of $\sim$65\% for the conversion yields of species with no detected IR features. 

\begin{figure*}
    \centering
    \includegraphics[width=13cm]{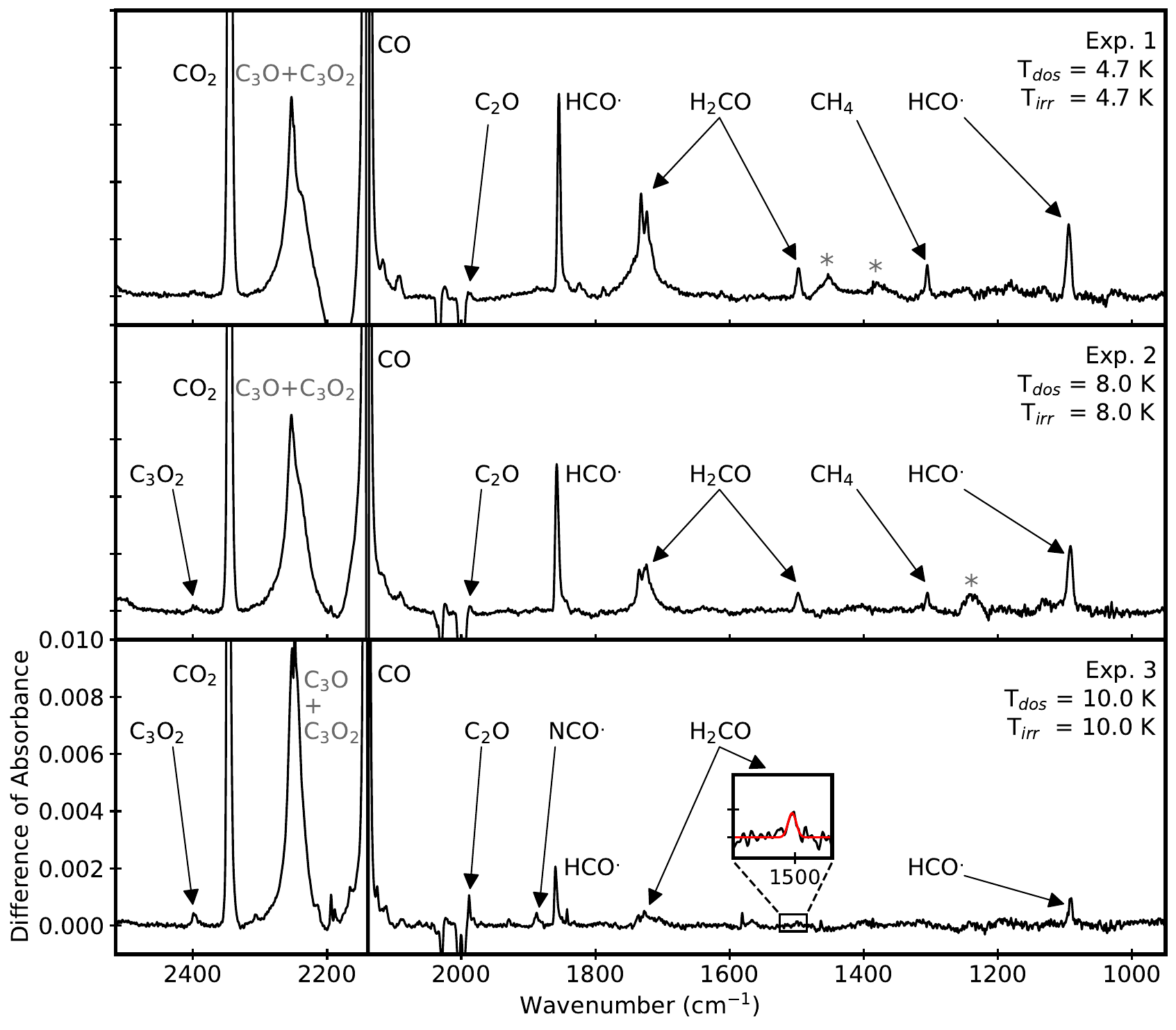}
    \caption{IR difference spectra obtained upon irradiation of a similar incident energy with 2 keV electrons of CO:$^{15}$N$_2$:H$_2$ ice samples in Experiments 1$-$3. 
    The IR difference spectra were baseline-corrected using a spline function. 
    IR band assignments are indicated in the panels.  
    Tentative assignments are indicated in grey.
    Grey asterisks indicate possible baseline artifacts that were not subtracted. 
    The inset in the bottom panel shows a zoom-in of the H$_2$CO feature observed at $\sim$1500 cm$^{-1}$ in the IR spectrum collected after irradiation (black), along with a Gaussian fit of the feature (red).}
    \label{fig:ir_series1}
\end{figure*}{}

{}

\begin{figure*}
    \centering
    \includegraphics[width=13cm]{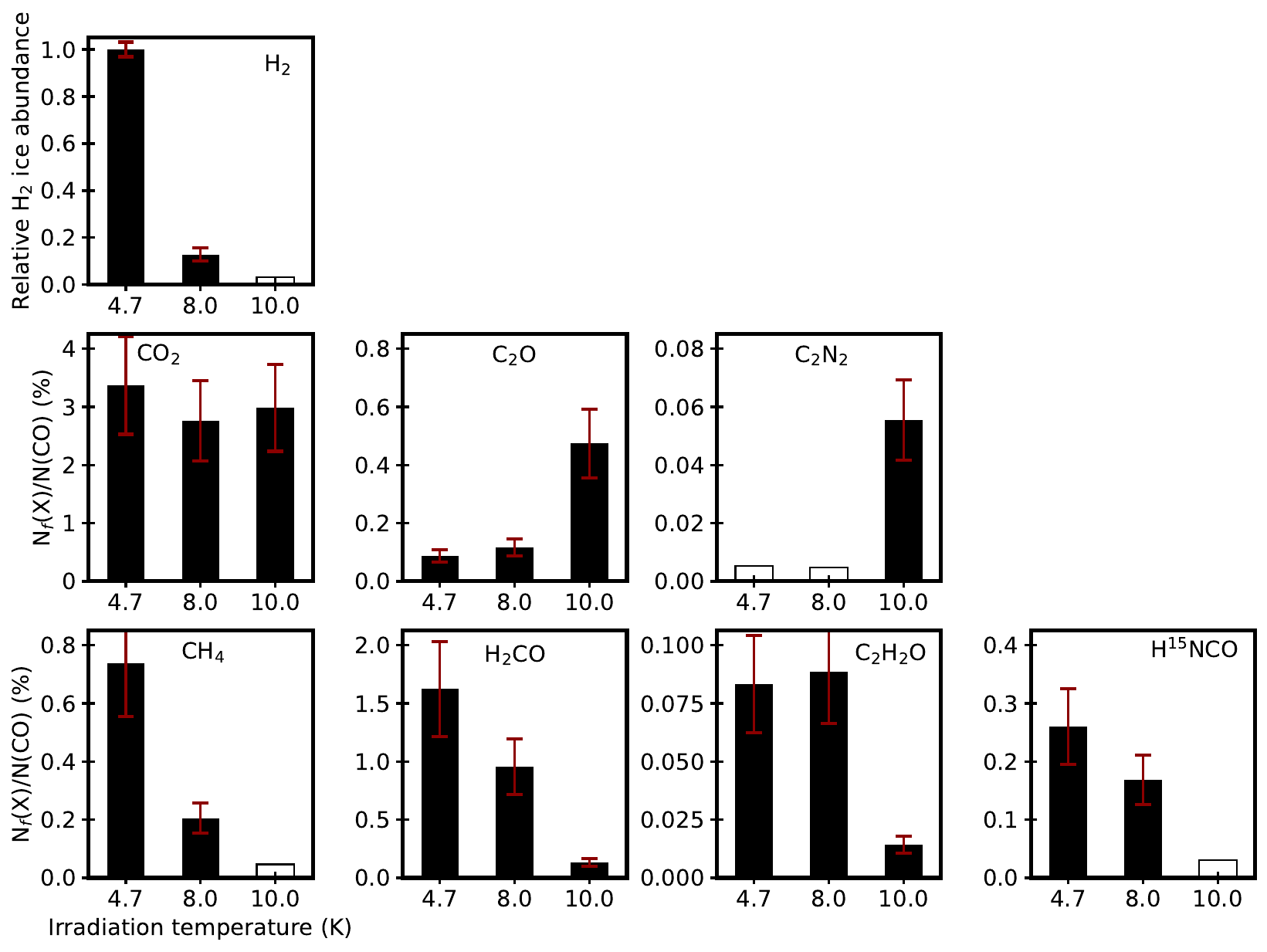}
    \vspace{-.5mm}
    \caption{Relative H$_2$ initial ice abundance with respect to Exp. 1 (top panel), and product yields with respect to the active CO ice column density of non-H-bearing (middle panels) and H-bearing (bottom panels) ice chemistry products in Experiments 1$-$3. 
    Empty bars indicate 3$\sigma$ upper limits. 
    Red errorbars represent the 25\% experimental uncertainty (except for the top panel, where they indicate the uncertainty of the H$_2$ IR feature Gaussian fit). 
    }
    \label{fig:yields_series1}
\end{figure*}{}

\section{Results}\label{sec:results}

As mentioned in Section \ref{sec:exp}, two series of experiments were performed. 
Even though the composition of the dosed gas mixture was similar in Experiments 1$-$6 (third column of Table \ref{table_exp}), the composition of the deposited ice samples  depended on the dosing temperature and the thermal evolution of the samples prior to irradiation. 
Section \ref{sec:h2} evaluates the initial ice composition in Experiments 1$-$6, and Sections \ref{sec:series1} and 
\ref{sec:series2} present the chemistry observed upon irradiation of the ice samples in Experiments 1$-$3 and 4$-$6, respectively.

\subsection{Initial ice composition}\label{sec:h2}

Fig. \ref{fig:ir_series1_initial} shows the IR spectra in the 4300$-$4100 cm$^{-1}$ range of the CO:$^{15}$N$_2$:H$_2$ ice samples collected prior to irradiation, but at the irradiation temperature in Experiments 1$-$3 (left panels) and 4$-$6 (right panels). 
The integrated absorbance of the CO IR feature detected at 4253 cm$^{-1}$ was used to calculate the initial CO ice column density, that ranged between 320 ML and 460 ML (fifth column of Table \ref{table_exp}).

Even though H$_2$ molecules are IR inactive due to their lack of a dipole moment, IR features can be induced when H$_2$ molecules are embedded in different matrices \citep{warren1980}. 
In particular, an IR feature corresponding to this species was observed in the 4145$-$4130 cm$^{-1}$ range in Experiments 1$-$6 \citep[Fig. \ref{fig:ir_series1_initial}, see also][]{loeffler10}.  
Table \ref{table_exp} lists the integrated absorbance of the H$_2$ IR feature calculated with a Gaussian fit. 
Unfortunately, the band strength of this feature is not reported in the literature, and the H$_2$ ice column density could not be calculated using Eq. \ref{eqn}. 
Instead, we estimated the H$_2$ ice abundance in Exp. 1 by assuming, as a first approximation, a 0.22 sticking coefficient for H$_2$ at 4.7 K \citep{german21}.  
We then used the relative H$_2$ IR integrated absorbances in Experiments 2$-$6 with respect to Exp. 1 to estimate the H$_2$ abundance in those other experiments (assuming the same H$_2$ IR band strength for the H$_2$ molecules embedded in the ice samples of Experiments 1$-$6).  
This resulted in an estimated H$_2$/CO ratio of $\sim$0.7 in Exp. 1. 
In Exp. 2, the H$_2$ IR integrated absorbance decreased by a factor of $\sim$10 compared to Exp. 1, resulting in an estimated H$_2$/CO ratio of $\sim$0.1. 
In Exp. 3, the H$_2$ IR feature was not detected (bottom left panel of Fig. \ref{fig:ir_series1_initial}), and only a $\le$0.03 upper limit could be estimated for the H$_2$/CO ratio (Table \ref{table_exp}). 
This could be explained by a decrease in the H$_2$ molecule sticking coefficient at higher temperatures, as previously reported in \citet{chuang18}.
On the other hand, ice samples grown at 6 K and warmed up to a higher temperature (Experiments 4$-$6) were able to entrap more H$_2$ molecules than ices grown at the same final temperature with the same dosing conditions. 
As a result, the H$_2$/CO ratio was $\sim$0.25 in the ice samples formed at 6 K and warmed up to 7 K and 9 K, 
while the ratio decreased to $\sim$0.1  in the ice formed at 6 K and warmed up to 15 K. 

Like H$_2$ molecules, N$_2$ molecules also present a weak IR feature at $\sim$2325 cm$^{-1}$ when they are embedded in a polar ice matrix. However, this feature is much weaker than the H$_2$ IR band \citep{loeffler06}, and was not detected in any of our experiments. 
Therefore, we assumed that the $^{15}$N$_2$/CO ice ratio was the same as in the gas mixture. 
The estimated initial ice composition in Experiments 1$-$6 is indicated in the fourth column of Table \ref{table_exp}. 

\begin{figure*}
    \centering
    \includegraphics[width=12cm]{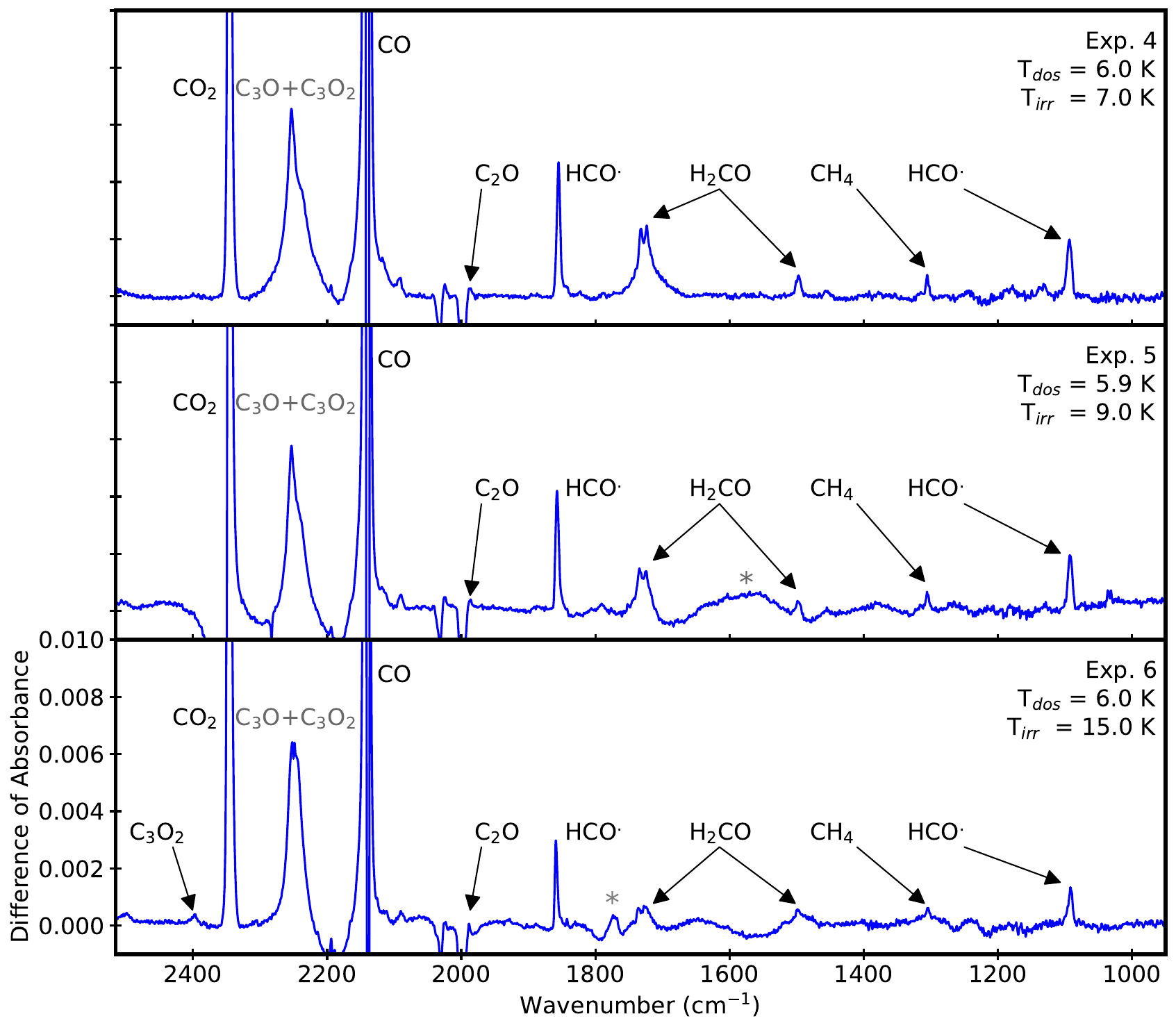}
    \caption{IR difference spectra obtained upon irradiation of a similar incident energy with 2 keV electrons of CO:$^{15}$N$_2$:H$_2$ ice samples in Experiments 4$-$6. 
    The IR difference spectra were baseline-corrected using a spline function with the \texttt{IDL} software. 
    IR band assignments are indicated in the panels.  
    Tentative assignments are indicated in grey.
    Grey asterisks indicate possible baseline artifacts that were not subtracted. 
    }
    \label{fig:ir_series2}
\end{figure*}{}

\begin{figure*}
    \centering
   \gridline{
    \fig{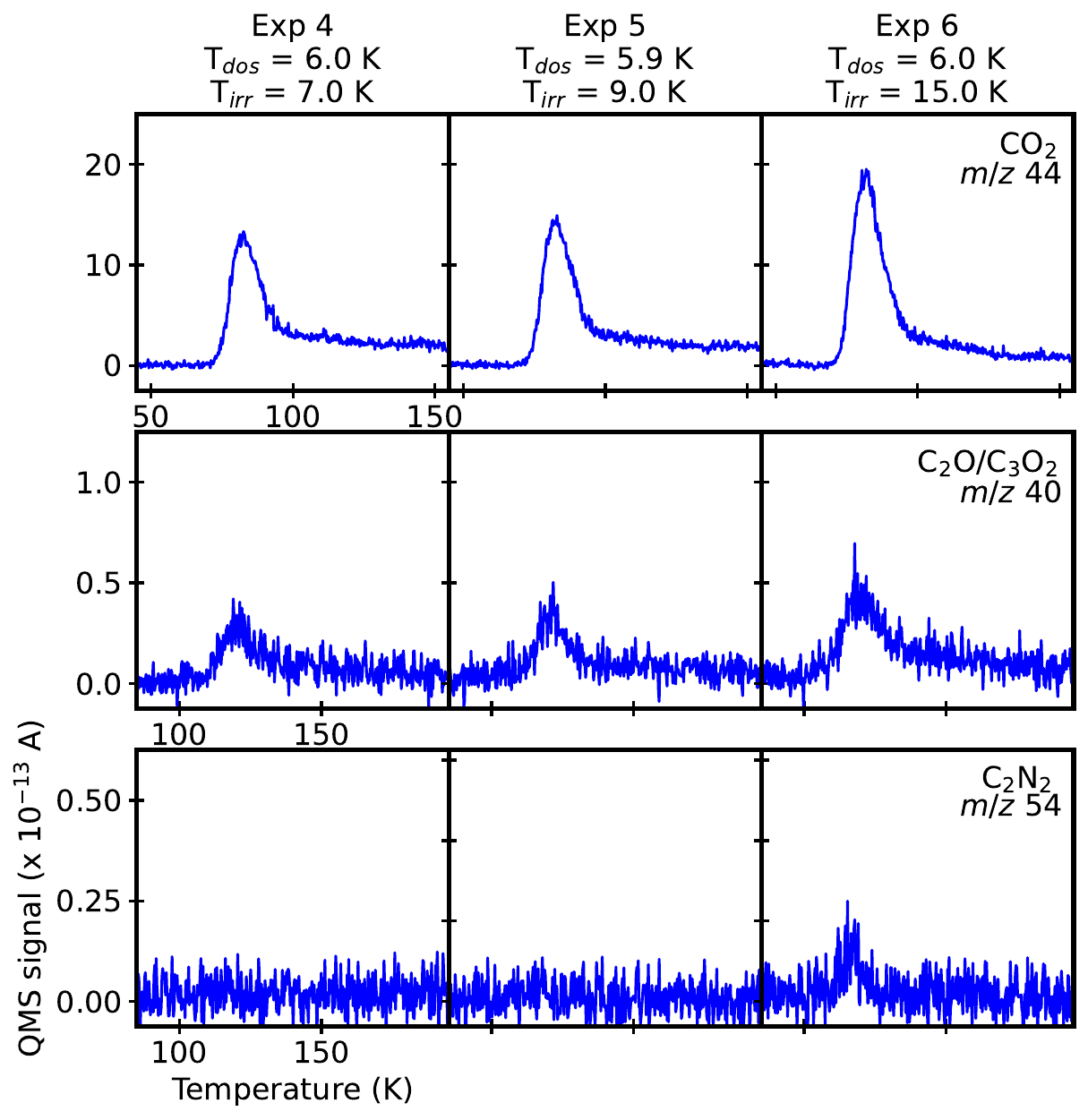}{0.49\textwidth}{}
    \fig{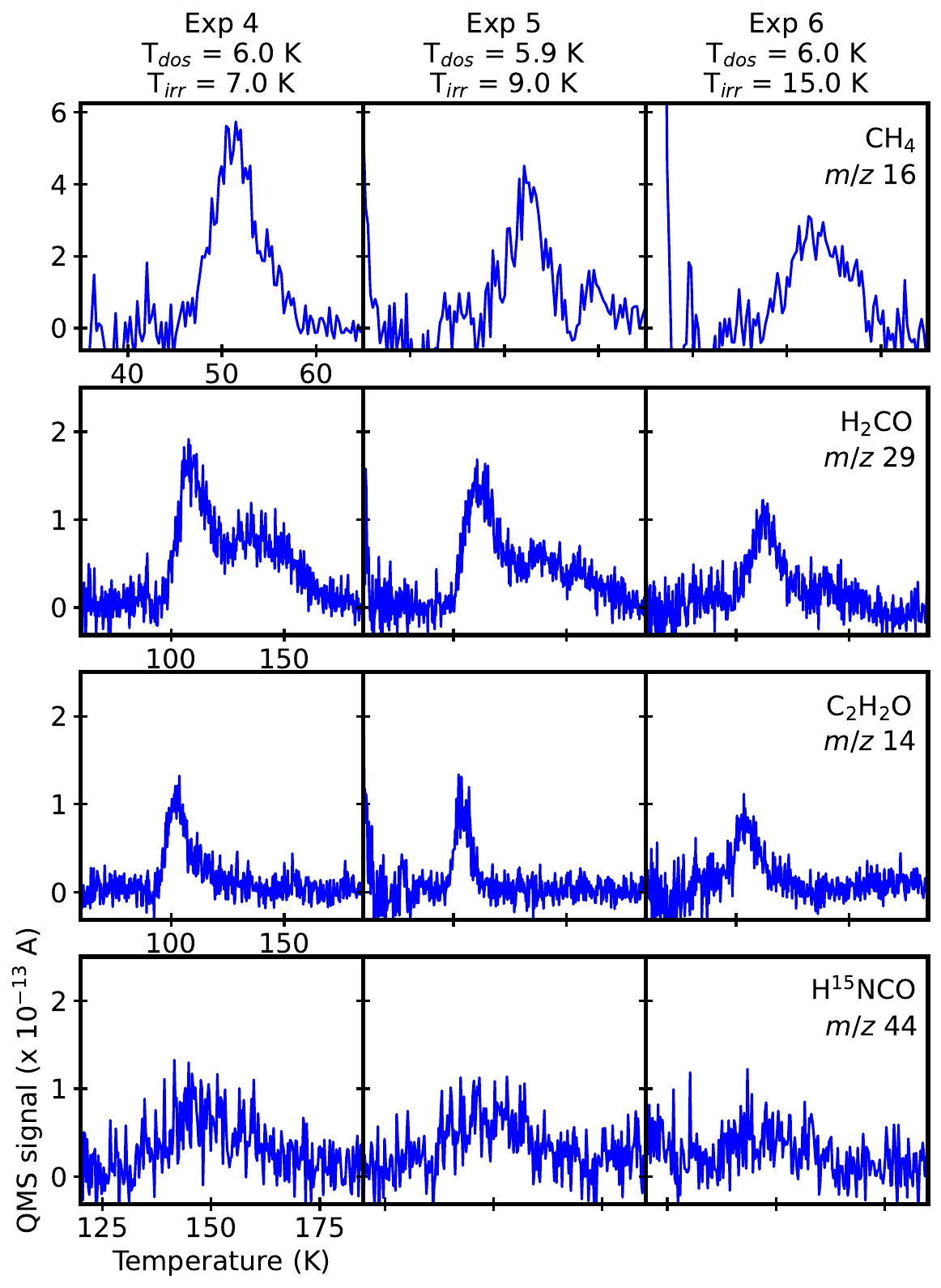}{0.475\textwidth}{}}
    \caption{TPD curves of the CO$_2$, C$_2$O/C$_3$O$_2$, and C$_2^{15}$N$_2$ main mass fragments (\textit{Left}, from top to bottom: $m/z$ 44, $m/z$ 40, and $m/z$ 54, respectively), and the CH$_4$, H$_2$CO, C$_2$H$_2$O, and H$^{15}$NCO  main mass fragments (\textit{Right}, from top to bottom: $m/z$ 16, $m/z$ 29, $m/z$ 14, and $m/z$ 44, respectively) during warm-up of CO:$^{15}$N$_2$:H$_2$ ice samples after irradiation of a similar incident energy with 2 keV electrons in Experiments 4$-$6 (from left to right).
    The TPD curves were baseline-corrected using a spline function with the \texttt{IDL} software. }
    \label{fig:tpd_series2}
\end{figure*}{}

\begin{figure*}
    \centering
    \includegraphics[width=13cm]{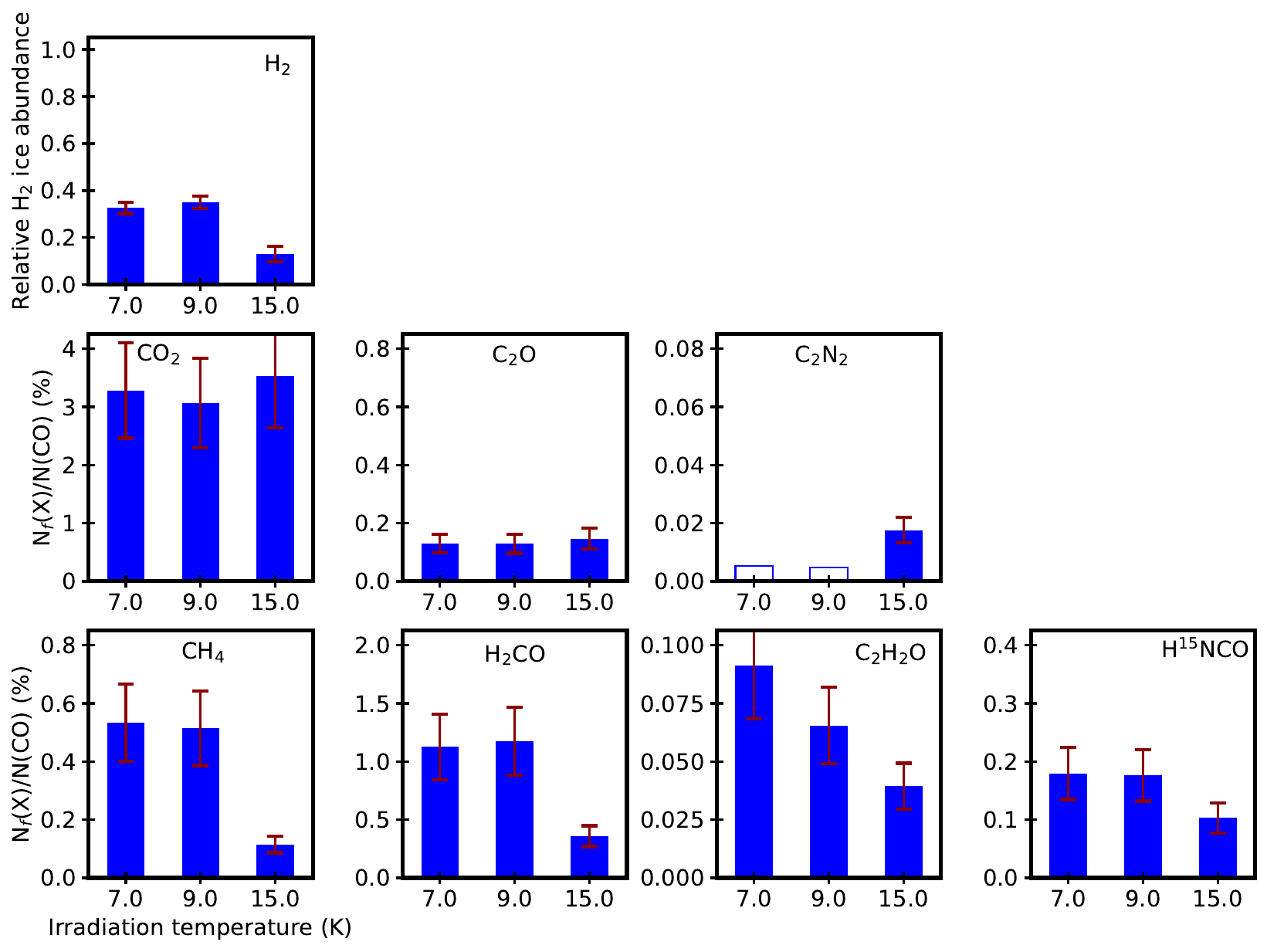}
    \vspace{-.5mm}
    \caption{Relative H$_2$ initial ice abundance with respect to Exp. 1 (top panel), and product yields with respect to the active CO ice column density of non-H-bearing (middle panels) and H-bearing (bottom panels) ice chemistry products in Experiments 4$-$6.   
    Empty bars indicate 3$\sigma$ upper limits. 
    Red errorbars represent the 25\% experimental uncertainty (except for the top panel, where they indicate the uncertainty of the H$_2$ IR feature Gaussian fit). }
    \label{fig:yields_series2}
\end{figure*}{}

\subsection{Organic chemistry in irradiated CO:$^{15}$N$_2$:H$_2$ ice samples grown at different temperatures}\label{sec:series1}

Figure \ref{fig:ir_series1} shows the IR difference spectra in the 2500$-$950 cm$^{-1}$ range of the CO:$^{15}$N$_2$:H$_2$ ice samples formed and irradiated at 4.7, 8.0, and 10.0 K  (Experiments 1$-$3, from top to bottom). 
The IR difference spectra were baseline-corrected using a spline function. 
However, some baseline artifacts could not be subtracted and are flagged with grey asterisks in Fig. \ref{fig:ir_series1}. 
IR features corresponding to the formation of  CO$_2$ (2348 cm$^{-1}$), C$_2$O (1989 cm$^{-1}$), and other carbon chain oxides such as C$_3$O$_2$ (2398 cm$^{-1}$ and $\sim$2245 cm$^{-1}$) and C$_3$O ($\sim$2250 cm$^{-1}$) were detected in all three experiments. 
We note that carbon chain oxide chemistry in CO-bearing ices was studied in detail in \citet{sicilia12}. 
IR features corresponding to H-bearing products were also detected in all experiments, including Exp. 3 in which the ice was deposited and irradiated at 10 K. 
This indicates that some H$_2$ molecules were present in the ice deposited at 10 K, even though the corresponding H$_2$ IR feature was not strong enough to be detected (bottom left panel of Fig. \ref{fig:ir_series1_initial}).  
In particular, two IR features corresponding to H$_2$CO were detected at $\sim$1725 and 1498 cm$^{-1}$ in the ice samples irradiated at 4.7, 8.0, and 10.0 K. 
The IR absorbance was significantly lower in the ice irradiated at 10.0 K (Fig. \ref{fig:ir_series1}, bottom panel).  
In addition, another two IR features corresponding to unreacted HCO$^.$ radicals were detected at $\sim$1850 and 1095 cm$^{-1}$. 
The absorbance of these features also decreased significantly in Exp. 3. 
An IR feature corresponding to CH$_4$ molecules (1305 cm$^{-1}$) was detected in the 4.7 and 8.0 K irradiation experiments (top and middle panels of Fig. \ref{fig:ir_series1}, respectively), but was not observed in Exp. 3.

The baseline-corrected TPD curves of CO$_2$, C$_2$O and C$_3$O$_2$, H$_2$CO, and CH$_4$ in Experiments 1$-$3 are shown in Fig. \ref{fig:tpd_series1}. 
%
We note that the main mass fragment of both, C$_2$O and C$_3$O$_2$ was $m/z$ 40.  The broad desorption feature of the $m/z$ 40 TPD curves probably presented contributions from both species. 
Unfortunately, the molecular mass of C$_3$O$_2$ was not monitored in Experiments 1$-$6, so its contribution to the $m/z$ 40 TPD curves could not be constrained. 
On the other hand, the C$_3$O mass spectrum was not reported in the NIST online database, and its molecular mass was not monitored either in Experiments 1$-$6. This, combined with the overlap of the C$_3$O IR feature with a C$_3$O$_2$ IR band made us consider the C$_3$O detection as tentative. 
The H$_2$CO TPD curves also presented broad desorption features, probably due to the formation of H$_2$CO dimers as explained in \citet{martin20}. In addition, 
CH$_4$ was relatively unconstrained from the TPD due to the contribution of CO fragmentation to $m/z$ 16. 
%
The TPD curves in Fig. \ref{fig:tpd_series1} show a significant increase in the formation of C$_2$O and C$_3$O$_2$ in the ice irradiated at 10.0 K (also observed in IR spectrum, bottom panel of Fig. \ref{fig:ir_series1}), paired with a decrease in the QMS signal of the H-bearing products H$_2$CO and CH$_4$. 

In addition, Fig. \ref{fig:tpd_series1} shows the TPD curves of three products 
whose formation could only be confirmed by the QMS during thermal desorption of the irradiated ice samples: C$_2^{15}$N$_2$, C$_2$H$_2$O, and H$^{15}$NCO. 
IR features 
corresponding to C$_2$H$_2$O and H$^{15}$NCO could not be unambiguously detected since their position overlapped with the CO and carbon chain oxide IR bands, respectively \citep{broekhuizen04,hudson20}; 
while the expected C$_2^{15}$N$_2$ IR integrated absorbance was below our detection limit \citep{hudson23}.
Formation of C$_2$H$_2$O was detected in Experiments 1$-$3 but, like H$_2$CO, it significantly decreased in the ice irradiated at 10.0 K. 
Formation of H$^{15}$NCO was only detected in the samples irradiated at 4.7 and 8.0 K, while formation of C$_2^{15}$N$_2$ was only observed in the ice irradiated at 10.0 K. 
We note that 
the assignment of all identified products was confirmed in \citet{martin20} after energetic processing of isotopically-labeled CO:N$_2$:H$_2$ ice samples. 
C$_2$N$_2$ was not produced in such ice samples, but the observed desorption temperature agreed with the estimated condensation temperature in Titan's atmosphere \citet{coustenis99}.

In order to quantify the differences in the chemistry of Experiments 1$-$3, Fig. \ref{fig:yields_series1} presents the conversion yields of the non-H-bearing products (middle panels) and H-bearing products (bottom panels) in the ices deposited and irradiated at 4.7, 8.0, and 10.0 K. 
In addition, the top panel of Fig. \ref{fig:yields_series1} shows the relative initial H$_2$ abundance of the ice samples in Experiments 1$-$3. 
The H$_2$ abundance in the ices deposited at 8.0 K and 10.0 K were 13\% and $\le$5\% of the value measured at 4.7 K, respectively.
The ice irradiation temperature along with the initial H$_2$ ice abundance affected the formation of the different products in different ways. 
CO$_2$ was the most abundant product in the three irradiated ice samples, with a conversion yield close to $\sim$3\% independent from the irradiation temperature and the H$_2$ abundance. 
%
Other non-H-bearing products, such as C$_2$O (selected as a representative of the carbon chain oxide chemistry) and C$_2^{15}$N$_2$ also presented similar conversion yields in the ice samples deposited and irradiated at 4.7 and 8.0 K.   
A factor of 4 and 10 increase, respectively, was observed in their conversion yields in the ice irradiated at 10.0 K, corresponding to the lowest measured H$_2$ ice abundance. 
On the other hand, the H-bearing products CH$_4$, H$_2$CO, C$_2$H$_2$O, and H$^{15}$NCO presented the opposite behavior, with a factor of $>$4 decrease in the yields of the ice irradiated at 10.0 K compared to the irradiation at lower temperatures. 
The observed anti-correlation in the yields of H-bearing and non-H-bearing molecules suggests that the formation of these two families of products constituted somewhat competing processes.
Similar yields (within the experimental uncertainties) were observed for the H-bearing products in the ices irradiated at 4.7 K and 8.0 K despite the factor of $\sim$10 difference in the H$_2$ abundance. 
The only exception was CH$_4$, that presented a factor of $\sim$3 decrease in Exp. 2 compared to Exp. 1. 
A possible explanation to the strongest dependence of the CH$_4$ yield with the H$_2$ abundance is that while the formation of H$_2$CO, C$_2$H$_2$O, and H$^{15}$NCO required only one H$_2$ molecule, CH$_4$ formation needed the participation of at least two H$_2$ molecules, making it more sensitive to variations in the H$_2$ abundance.


\subsection{Organic chemistry in CO:$^{15}$N$_2$:H$_2$ ice samples grown at 6 K and irradiated at 6$-$15 K}\label{sec:series2}

Fig. \ref{fig:ir_series2} shows the baseline-corrected IR difference spectra of the CO:$^{15}$N$_2$:H$_2$ ice samples formed at $\sim$6 K and irradiated at 7.0, 9.0, and 15.0 K (Experiments 4$-$6, from top to bottom).  
As in Experiments 1$-$3, IR features corresponding to the non-H-bearing ice chemistry products CO$_2$, C$_2$O, C$_3$O$_2$, and C$_3$O were detected in the 2400$-$1990 cm$^{-1}$ range in the three experiments (Fig. \ref{fig:ir_series2}). 
On the other hand, the decrease in the IR absorbance of the features corresponding to the H-bearing products H$_2$CO, HCO$^{.}$ and CH$_4$ for increasing irradiation temperatures was not as significant as that observed in Experiments 1$-$3 (Sect. \ref{sec:series1}). 
In particular, the IR feature corresponding to CH$_4$ was observed in all three experiments.
This can be explained by the higher abundances of H$_2$ measured in the samples irradiated at 9 K and 15 K when the ice was initially deposited at 6 K, compared to the ices deposited at 8 K and 10 K (Sect. \ref{sec:h2})

The baseline-corrected TPD curves are shown in Fig. \ref{fig:tpd_series2}. 
The TPD results were consistent with the analysis of the IR difference spectra. 
We also observed a similar trend for the organics H$_2$CO, C$_2$H$_2$O, and H$^{15}$NCO, that were detected in all three experiments, and did not present a decrease in their QMS signals for increasing irradiation temperatures as significant as in Experiments 1$-$3. 

Fig. \ref{fig:yields_series2} presents the conversion yields of non-H-bearing (middle panels) and H-bearing (bottom panels) products, along with the relative H$_2$ initial ice abundance (top panel) in Experiments 4$-$6. 
The estimated H$_2$ abundance in the samples deposited at 6 K and warmed up to 7 K and 9 K was $\sim$35\% of the value at 4.7 K. This abundance decreased down to $\sim$13\% of the 4.7 K value when the ice was warmed up to 15 K. 
The product conversion yields were the same (within errors) in the samples irradiated at 7 K and 9 K (Experiments 4 and 5), that presented a similar H$_2$ abundance. 
These yields were also similar to those measured in the samples deposited and irradiated at 4.7 K and 8.0 K (Experiments 1 and 2), with the exception of CH$_4$, that presented a lower yield in Exp. 2 (Sect. \ref{sec:series1}). 
The conversion yields of the H-bearing products (CH$_4$, H$_2$CO, C$_2$H$_2$O, and H$^{15}$NCO) decreased by a factor of 1.7$-$4.5 in the ice irradiated at 15 K (Exp. 6).  This sample presented a factor of $\sim$3 lower H$_2$ abundance than in Experiments 4 and 5. However, this decrease did not prevent the detection of CH$_4$ and H$^{15}$NCO in Exp. 6, in contrast to their non-detection in Exp. 3. 
As explained above, this was probably due the higher H$_2$ abundance in the sample deposited at 6 K and irradiated at 15 K compared to the ice deposited and irradiated at 10 K. 
Regarding the non-H-bearing products, the conversion yields of CO$_2$ and C$_2$O were the same (within errors) in Experiments 4$-$6, while formation of C$_2^{15}$N$_2$ was only detected in the ice irradiated at 15.0 K.

\section{Discussion}\label{sec:disc}

\subsection{Effect of the temperature and the H$_2$ ice abundance in the organic chemistry of the CO-rich ice layer}

The top panel in Fig. \ref{fig:yields_series1} shows that the abundance of H$_2$ in CO-rich ices is very sensitive to the ice formation temperature, as already suggested in \citet{chuang18}.
In any case, the results presented in Sections \ref{sec:series1} and \ref{sec:series2} indicate that the chemistry induced by electron irradiation of H$_2$-bearing, CO-rich ices is somewhat robust for ices with H$_2$ abundances spanning one order of magnitude in the 4$-$9 K temperature range (Experiments 1, 2, 4, and 5). 
Focusing on the organic chemistry, similar conversion yields were measured for H$_2$CO, C$_2$H$_2$O, and H$^{15}$NCO in ice samples irradiated at T$_{irr}$ = 4.7 K with an estimated H$_2$/CO ratio of $\sim$0.7 (Exp. 1), T$_{irr}$ = 7.0 K with H$_2$/CO $\sim$ 0.25 (Exp. 4), T$_{irr}$ = 8.0 K with H$_2$/CO $\sim$ 0.1 (Exp. 2), and T$_{irr}$ = 9.0 K with H$_2$/CO $\sim$ 0.25 (Exp. 5). 
This H-rich chemistry gave way to a more H-poor chemistry in the ice irradiated at 10.0 K with a H$_2$/CO ratio $\le$0.03 (Exp. 3). This experiment presented a decrease of a factor of $>$5 in the conversion yields of H$_2$CO, C$_2$H$_2$O, and H$^{15}$NCO.
The H-poor chemistry was also characterized by higher conversion yields of some non-H-bearing products such as C$_2$O and C$_2^{15}$N$_2$. The latter was not detected in ice samples with H-rich chemistry. 
An intermediate case was observed in Exp. 6, with an irradiation temperature of 15 K and a H$_2$/CO ratio of $\sim$0.1. In this case, the decrease in the organic conversion yields was not as significant as in Exp. 3, and all H-bearing products were detected along with C$_2^{15}$N$_2$. 


In addition, we could evaluate the effect of the ice irradiation temperature independently from the H$_2$ abundance by comparing Experiments 2 and 6, that presented similar H$_2$ abundances but were irradiated at a factor of $\sim$2 different temperatures (8 K and 15 K, respectively). 
%
%
We observed a factor of 2$-$3 decrease in the yields of the organics H$_2$CO, C$_2$H$_2$O and H$^{15}$NCO for the ice irradiated at 15 K (Exp. 6), which we attributed to shorter H-atom residence times following H$_2$ dissociation compared to the 8 K experiment (Exp. 2).

\subsection{Relative contribution of the H$_2$-bearing, CO-rich ice layer to the interstellar COM chemistry}

\citet{martin20} discussed in depth the potential contribution of the energetic processing of the H$_2$-bearing, CO-rich interstellar ice layer to the formation of COMs in the interior of dense clouds. 
Following those results, the laboratory experiments presented in this work show that the formation of four H-bearing species is observed when irradiating CO:N$_2$:H$_2$ ice samples with dissociative radiation under conditions relevant to the interior of dense clouds: CH$_4$, H$_2$CO, C$_2$H$_2$O, and HNCO. 
Three of these species (H$_2$CO, C$_2$H$_2$O, and HNCO) are to some extent related to the complex organic chemistry, including HNCO, the simplest molecule to contain the amide bond. 

H$_2$CO has been recently confirmed to be present in interstellar ice mantles through JWST/MIRI observations toward the IRAS 15398-3359 class 0 protostar \citep{yang22}. 
Previous observations estimated an abundance of 2$-$7\% with respect to H$_2$O ice in different environments \citep{boogert15}. 
However, it is currently unclear whether this species is present in the CO-rich and/or the H$_2$O-rich ice layers \citep[see][and references therein]{qasim19b}. 
H$_2$CO has been proposed to form in the CO-rich layer through successive H-atom additions to CO molecules \citep[see, e.g.,][]{watanabe02,fuchs09}. 
However, in the interior of dense clouds, where temperatures could decrease down to $<$10 K, the H/H$_2$ ratio decreases rapidly with density \citep{boogert15}. 
Sections \ref{sec:series1} and \ref{sec:series2} show that at temperatures $<$10 K the H$_2$CO conversion yield upon electron irradiation of H$_2$-bearing, CO-rich ices is $>$3 times higher than the yield measured at higher temperatures. 
This, combined with a high H$_2$/H abundance ratio, could increase the relative contribution of the H$_2$CO formation pathway presented in this work compared to the H-atom addition pathway in the coldest regions of the ISM. 

C$_2$H$_2$O has been observed in the gas phase of translucent clouds \citep{turner99}, pointing to a solid-phase formation of this species contemporary to the formation of the H$_2$O-rich ice layer, followed by subsequent desorption to the gas phase \citep{chuang20}. 
The observed increase by a factor of up to $\sim$6 in the conversion yield of this species upon irradiation of a H$_2$-bearing, CO-rich ices at temperatures lower than 10 K (Sections  \ref{sec:series1} and \ref{sec:series2}) could also boost the relative contribution of this formation pathway compared to the one proposed in the literature \citep{chuang20} in the cold ISM.

Contribution of solid-phase formation pathways have previously been invoked to explain the observed HNCO gas-phase abundances \citep{tideswell10,quan10,quenard18}. 
However, it is not obvious that the different solid-phase formation mechanisms proposed in the literature can fully account for the observed abundances of HNCO in different regions of the ISM \citep[see][and references therein]{martin20}. 
Therefore, the energetic processing of ice mantles containing CO, N$_2$, and H$_2$ could represent a significant contribution to the formation of this species in the ISM. 


\section{Conclusions}\label{sec:conc}

\begin{enumerate} 

    \item The ice abundance of H$_2$ was 
    very sensitive to the ice formation temperature.  In our experiments, compared to a CO:$^{15}$N$_2$:H$_2$ ice sample deposited at T$_{dos}$ = 4.7 K, the estimated H$_2$ ice abundance decreased by a factor of $\sim$3 for T$_{dos}$ = 6 K, a factor of $\sim$8 for  T$_{dos}$ = 8 K, and by more than a factor of 20 for T$_{dos}$ = 10 K. 

    \item CO-rich ice samples formed at 6 K and warmed up to higher temperatures were able to entrap a larger amount of H$_2$ molecules within the ice matrix than ices formed at higher temperatures. As a result, the estimated H$_2$ abundance at 15 K for an ice deposited at 6 K was more than $\sim$3 times higher than the abundance of an ice deposited at 10 K. 
    
    \item The presence of H$_2$ molecules in CO-rich ices enabled the formation of organic molecules related to COMs (such as H$_2$CO, C$_2$H$_2$O, and H$^{15}$NCO) induced by energetic processing of the ice. 
    This organic chemistry took place to a similar extent in ices spanning an order of magnitude in the estimated H$_2$ abundance, and was enhanced by a factor of $\sim$5 at temperatures $<$10 K compared to warmer ices. The decrease in warmer ices could be due to a combination of lower initial H$_2$ abundances and shorter residence times of the H atoms resulting from the dissociation of the H$_2$ molecules. 

    \item Further experiments comparing the relative conversion yields of the different solid-phase formation pathways of H$_2$CO, C$_2$H$_2$O, and HNCO, combined with astrochemical simulations should be performed to address the relative contribution of the energetic processing of the H$_2$-bearing, CO-rich ice layer to the interstellar budget of these organic species.

\end{enumerate}

\begin{acknowledgments}
This work was supported by a grant from the Simons Foundation (686302, K\"O) and an award from the Simons Foundation (321183FY19, K\"O). 
The project was also supported in part by the National Science Foundation REU and Department of Defense ASSURE programs under NSF Grant no. AST-2050813, and by the Smithsonian Institution. 
\end{acknowledgments}

%






\appendix

\section{Ice sample preparation}

\subsection{Gas mixture composition}\label{sec:gas_composition_app}

\begin{figure*}
    \centering
    \includegraphics[width=11cm]{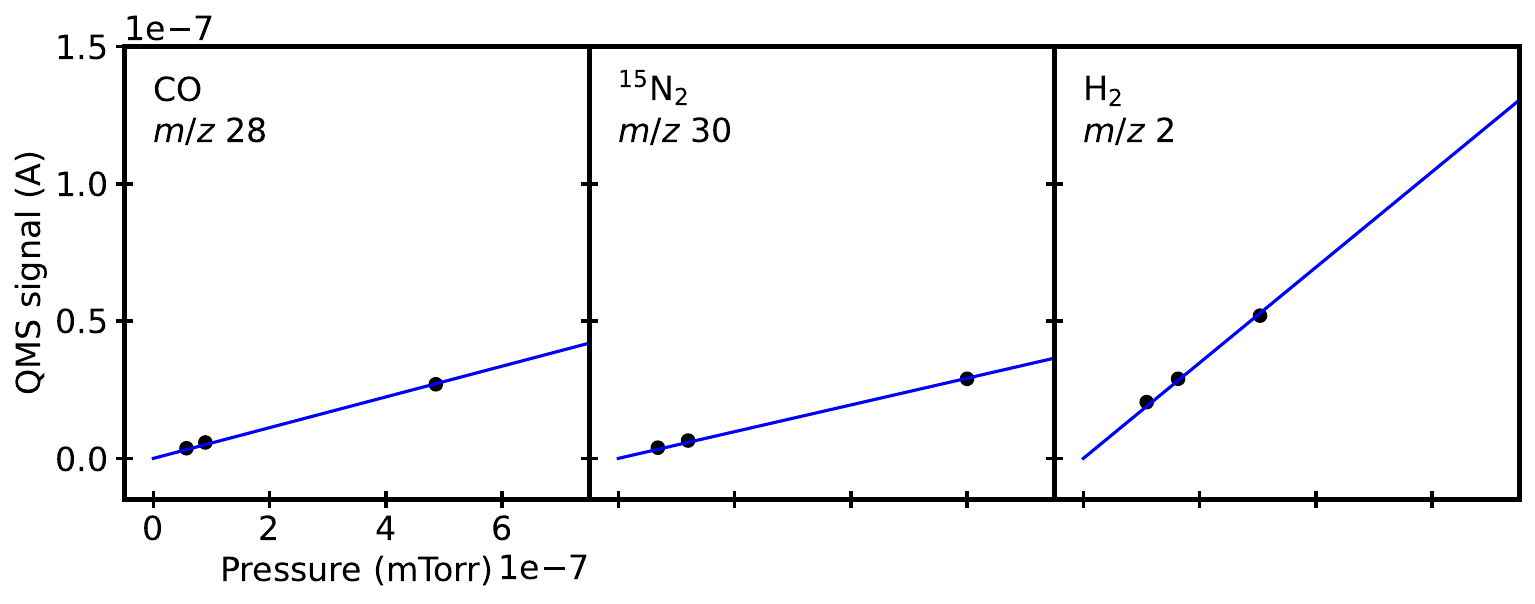}
    \caption{Linear fit (solid blue line) to the measured QMS signal of the main mass fragments of CO ($m/z$ 28, left panel), $^{15}$N$_2$ ($m/z$ 30, middle panel), and H$_2$ ($m/z$ 2, right panel) at different pressures of the pure gases in SPACE TIGER (black dots).}
    \label{fig:qms_calib}
\end{figure*}{}

The gas mixture composition ratio indicated in the third column of Table \ref{table_exp} was measured in the UHV chamber before ice deposition with the help of the QMS (Sect. \ref{sec:tpd}). 
To that purpose, a conversion factor for each species was previously calculated in order to transform the measured QMS signal of the main mass fragment corresponding to CO ($m/z$ 28 amu), $^{15}$N$_2$ ($m/z$ 30 amu), and H$_2$ ($m/z$ 2 amu) into partial pressures. 
This conversion factor was calculated by introducing different pressures of the pure gases in the UHV chamber, 
and measuring the QMS signal of the species main mass fragment. 
We note that the pressure read by the baratron gauge attached to the UHV chamber was corrected by the relative probability of ionization with respect to N$_2$, as indicated by the manufacturer. 
Fig. \ref{fig:qms_calib} shows the linear dependence of the QMS signal with the corrected gas pressure for CO, $^{15}$N$_2$, and H$_2$ molecules.

\subsection{QMS calibration}\label{sec:qms_calib_app}

\begin{figure}
    \centering
    \includegraphics[width=5cm]{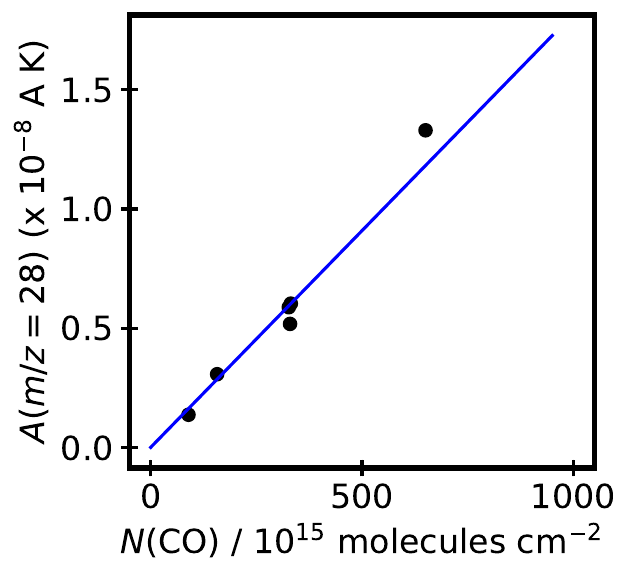}
    \caption{Linear relation (solid blue line) between the area under the TPD curve of a pure CO ice in SPACE TIGER ($A(m/z$ 28)) and the CO ice column density $N$(CO), extracted from a series of seven calibration experiments (black dots).}
    \label{fig:CO_calib}
\end{figure}{}

\begin{figure}
    \centering
    \includegraphics[width=5cm]{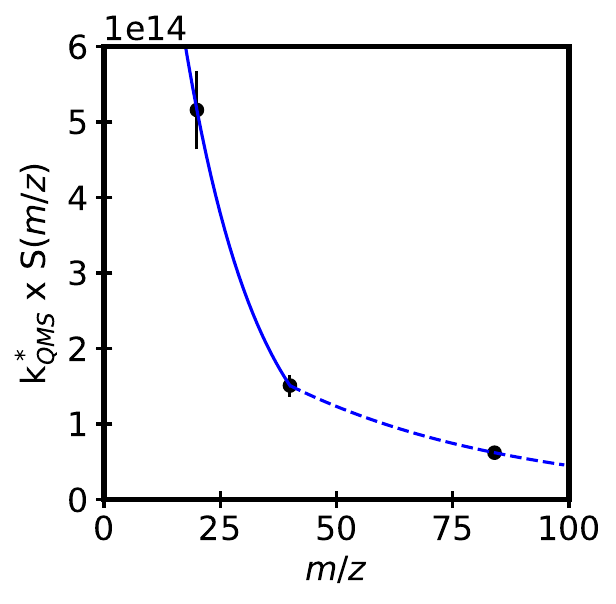}
    \caption{Exponential dependence of  $k^*_{QMS} \cdot S(m/z)$ with $m/z$ in SPACE TIGER QMS in the $m/z <$ 40 (solid blue line) and $m/z >$ 40 (dahsed blue line) range,  extracted from a series of calibration experiments using three different noble gases (black dots).}
    \label{fig:qms_sens}
\end{figure}{}

In order to estimate ice column densities and conversion yields for the identified species in Experiments 1$-$6 using the corresponding QMS signal measured during TPD of the irradiated ice samples, a series of additional calibration experiments were performed as part of this work. 

The area under the TPD curve of any mass fragment $m/z$ corresponding to species $X$ is proportional to that species ice column density, and can be calculated with the equation: 

\begin{equation}
    A(m/z) = k_{QMS} \cdot N(X) \cdot \sigma^{+}(X)  \cdot I_F(z) \cdot F_F(m) \cdot S(m/z) \label{eq_qms}
\end{equation}

where $A(m/z)$ is the area under the TPD curve, $k_{QMS}$ is a proportionality constant, $N(X)$ is the species ice column density, $\sigma^{+}(X)$ is the electron-impact ionization cross-section for the corresponding electron energy of the QMS, $I_F(z)$ is the fraction of ionized molecules with charge $z$, $F_F(m)$ is the fraction of molecules leading to a fragment of mass $m$ in the QMS, and $S(m/z)$ is the sensitivity of the QMS to the mass fragment $m/z$ \citep[see][and references therein]{martin15}.  

In the case of the $m/z$ 28 molecular fragment corresponding to CO molecules, Eq. \ref{eq_qms} could be written as: 

\begin{equation}
    A(m/z \: 28) = k_{QMS} \cdot N(CO) \cdot \sigma^{+}(CO)  \cdot I_F(1) \cdot F_F(28) \cdot S(28) = k(CO) \cdot N(CO) \label{eq_qms_2}
\end{equation}

where $k(CO)$ is 

\begin{equation}
    k_{CO} = \frac{A(m/z \: 28)}{N(CO)} = k_{QMS} \cdot \sigma^{+}(CO) \cdot I_F(1) \cdot F_F(28) \cdot S(28)
\end{equation}{}

We used a series of six calibration experiments to extract the proportionality constant $k_{CO}$. 
The calibration experiments consisted in the formation of pure CO ice samples at $\sim$4 K on top of an IR transparent CsI susbtrate that allowed the estimation of the CO ice column density $N(CO)$ from the IR spectra collected in transmission mode. 
The 2139 cm$^{-1}$ IR band was numerically integrated using the composite Simpson's rule (\texttt{integrate.simps} in SciPy)), and the column density was calculated with Eq. \ref{eqn} and the corresponding band strength listed in Table \ref{table_band}.
After deposition, we applied a heating rate of 2 K min$^{-1}$ to the ice sample until complete sublimation was achieved. 
The area under the CO $m/z 28$ TPD curve was also calculated with the \texttt{integrate.simps} function in SciPy). 
Fig. \ref{fig:CO_calib} shows the linear fit performed with the \texttt{curve$\_$fit} function in SciPy) to the six experimental data points. 
The proportionality constant $k_{CO}$ was found to be (1.82 $\pm$ 0.02) $\times$ 10$^{-11}$ A K ML$^{-1}$. 
We assumed a 20\% uncertainty due to the 20\% uncertainty of the CO IR band strength \citep{dhendecourt86}.

For species other than CO, the $N(X)$ column density in Experiments 1$-$6 could be calculated from the area under the corresponding TPD curve ($A(m/z)$) with the following equation: 

\begin{equation} 
N(X) = \frac{A(m/z)}{k_{CO}} \cdot \frac{\sigma^+(CO)}{\sigma^+(X)} \cdot \frac{I_F(CO^+)}{I_F(z)}
\cdot \frac{F_F(28)}{F_F(X)} \cdot \frac{S(28)}{S(m/z)},  \label{eqn_qms}
\end{equation}

using the parameters listed in Table \ref{table_qms}.

The calibration of the QMS sensitivity $S(m/z)$ was performed by introducing three noble gases into the UHV chamber at room temperature: Kr (gas, X, X\%), Ar (gas, X, X\%), and Ne (gas, X, X\%). 
The QMS ion current ($I(m/z)$) corresponding to the atomic masses of Kr ($m/z = 84$), Ar ($m/z 40$), and Ne ($m/z 20$) was proportional to the measured pressure $P(X)$, and could be calculated with an equation similar to Eq. \ref{eq_qms} but with a different proportionality constant $k^*_{QMS}$: 

\begin{equation}
    I(m/z) = k^*_{QMS} \cdot P(X) \cdot  \sigma^{+}(X)  \cdot I_F(z) \cdot Is_F(m) \cdot S(m/z). \label{eq_qms_noble}
\end{equation}

Note that the $F_F$ factor was not taken into account since atomic species cannot fragment in the QMS. Instead, an isotopic fraction $Is_F$ was included in order to only take into account the contribution of the isotopolog of mass $m$ to the total pressure $P(X)$ measured in the chamber. 

For every noble gas, the corresponding $k^*_{QMS} \cdot S(m/z)$ was calculated using the parameters listed in Table \ref{table_qms_noble} (note that the ratio $S(28)/S(m/z)$ used in Eq. \ref{eqn_qms} is equivalent to $k^*_{QMS} \cdot S(28)$ / $k^*_{QMS} \cdot S(m/z)$). 
Fig. \ref{fig:qms_sens} shows the sensitivity curve of SPACE TIGER QMS. We used the \texttt{curve$\_$fit} function in SciPy to describe the sensitivity for fragments in the $m/z <$ 40 and $m/z >$ 40 ranges with two exponential curves \citep[using the same expression as in][]{martin15}: 

\begin{equation}
    k^*_{QMS} \cdot S(m/z) = 1.56 \times 10^{15} \cdot e^{\frac{-m/z}{17.14}} (m/z < 40)
\end{equation}

\begin{equation}
    k^*_{QMS} \cdot S(m/z) = 3.39 \times 10^{14} \cdot e^{\frac{-m/z}{49.52}} (m/z > 40)
\end{equation}

\subsection{CO 4253 cm$^{-1}$ IR band strength in transmission and reflection-absorption mode}\label{sec:ir_app}

In addition to the estimation of $k_{CO}$, we used the calibration experiments presented in \ref{sec:qms_calib_app} to calculate the band strength of the CO 4253 cm$^{-1}$ IR feature in transmission mode. 
As explained above, we used the integrated absorbance of the 2139 cm$^{-1}$ IR feature (with a known band strength listed in Table \ref{table_band}) to calculate the CO ice column density in the calibration experiments. 
The 4253 cm$^{-1}$ IR feature was also numerically integrated using \texttt{integrate.simps} in SciPy), and the corresponding value of the band strength was calculated using Eq. \ref{eqn}. 
The average band strength calculated from the six calibration experiments was (7.76 $\pm$ 0.24) $\times$ 10$^{-20}$ cm molecule$^{-1}$. 
A 20\% uncertainty was assumed for this value, due to the 20\% uncertainty assumed in the CO 2139 cm$^{-1}$ IR feature band strength used to calculate the CO ice column density in the calibration experiments.  
We note that the 1$\sigma$ standard deviation in the band strength calculated for the different calibration experiments was lower than 10\%, and was considered negligible compared to the 20\% uncertainty in the estimation of the CO ice column density mentioned above. 

A similar procedure was followed for the estimation of the band strength of this feature in reflection-absorption mode. 
In this case, the CO ice column density after irradiation of the ice samples in Experiments 1$-$6 (Table \ref{table_exp}) was estimated from the area under the $m/z$ 28 TPD curve using the $k_{CO}$ proportionality constant derived in \ref{sec:qms_calib_app}. 
The 4253 cm$^{-1}$ IR feature observed in the IR spectra collected after irradiation was again numerically integrated, and the corresponding value of the band strength was calculated using Eq. \ref{eqn}, leading to an average band strength of (1.8 $\pm$ 0.4) $\times$ 10$^{-19}$ cm molecule$^{-1}$. 
A total uncertainty of 30\% was assumed for the band strength of this feature in reflection-absorption mode, due to the 22\% 1$\sigma$ standard deviation of this parameter from one experiment to another, and the 20\% uncertainty in the proportionality constant $k_{CO}$ used for the estimation of the CO ice column density.

\begin{deluxetable*}{cccccccc}
\tablecaption{Parameters used in Eq. \ref{eqn_qms} to estimate ice column densities from integrated QMS signals during TPD of the irradiated ice samples.\label{table_qms}}
\tablehead{
\colhead{Factor} & \colhead{CO} & \colhead{CO$_2$} & \colhead{C$_2$O}  &  \colhead{CH$_4$} & \colhead{C$_2^{15}$N$_2$} & \colhead{C$_2$H$_2$O} & \colhead{H$^{15}$NCO}}
\startdata
monitored $m/z$ & 28 & 44 & 40 & 16 & 54 & 14 & 44 \\
$\sigma^+(X)$ (\AA$^2$)$^b$ & 2.52 & 3.52 & 3.5$^c$ & 3.52 & 3.5$^c$ & 3.5$^c$ & 3.5$^c$\\ 
$I_F(z)^d$ & 1 & 1 & 1 & 1 & 1 & 1 & 1 \\
$F_F(m)^e$ & 0.89 & 0.64 & 0.5$^c$ & 0.40 & 0.95$^b$ & 0.35$^b$ & 0.5$^c$\\
$k^*_{QMS} \cdot S(m/z)$ ($\times$ 10$^{14}$ A mtorr$^{-1}$ \AA$^{-2}$)$^f$ & 3.04 & 1.39 & 1.51 & 6.88 & 1.14 & 6.88 & 1.39\\
\hline
\enddata
\tablecomments{
$^b$ Extracted from 
NIST database, unless otherwise indicated. 
$^c$Assumed as a first approximation. 
$^d$We assumed as a first approximation that double ionization of the molecules did not take place in the QMS.
$^e$Calculated empirically for the QMS in SPACE TIGER unless otherwise indicated. 
$^f$This work (Appendix \ref{sec:qms_calib_app})
}
\end{deluxetable*}{}

\begin{deluxetable*}{cccc}
\tablecaption{Parameters used in Eq. \ref{eq_qms_noble} to estimate the dependence of $k^*_{QMS} \cdot S(m/z)$ with $m/z$.\label{table_qms_noble}}
\tablehead{
\colhead{Factor} & \colhead{Kr} & \colhead{Ar} & \colhead{Ne}}
\startdata
$\sigma^{+}(X)$ (\AA$^2$)$^a$ & 3.450 & 2.520 & 0.475 \\ 
$I_F(z)^a$ & 0.8029 & 0.8909 & 0.9998 \\
$Is_F(m)$ & 0.9$^b$ & 0.9995$^b$ & 0.905$^c$\\
\hline
\enddata
\tablecomments{
$^a$From \citet{rejoub02}
$^b$From the provider specifications
$^c$From \citet{laeter03}}
\end{deluxetable*}{}




\end{document}